\documentclass[conference]{IEEEtran}
\IEEEoverridecommandlockouts
\usepackage{cite}
\usepackage{amsmath,amssymb,amsfonts}
\usepackage{algorithmic}
\usepackage{graphicx}
\usepackage{textcomp}
\usepackage{soul}
\def\BibTeX{{\rm B\kern-.05em{\sc i\kern-.025em b}\kern-.08em
    T\kern-.1667em\lower.7ex\hbox{E}\kern-.125emX}}
\begin{document}

\title{Real-time non-intrusive depth estimation of buried radioactive wastes based on approximate three-dimensional relative attenuation model \\
\thanks{Funded by the Engineering and Physical Sciences Research Council, UK and the National Decommissioning Authority, UK}
}

\author{\IEEEauthorblockN{Ikechukwu K. Ukaegbu}
\IEEEauthorblockA{\textit{Engineering Department} \\
\textit{Lancaster University}\\
Lancaster, LA1 4YW, UK \\
i.ukaegbu@lancaster.ac.uk}
\and
\IEEEauthorblockN{Kelum A. A. Gamage}
\IEEEauthorblockA{\textit{School of Engineering} \\
\textit{University of Glasgow}\\
Glasgow, G12 8QQ, UK \\
kelum.gamage@glasgow.ac.uk}
}

\maketitle

\begin{abstract}
A new method for non-intrusive estimation of the depth of buried radioactive waste have been developed. The method is based on an approximate three-dimensional relative attenuation model that exploits the variation in the intensities of the  radiation image obtained on the surface of the material in which the radioactive source is buried. Experimental results using an organic liquid scintillator detector showed that the method is able to estimate the depth of a 329 kBq Cs-137 radioactive point source buried up to 12 cm in sand.
\end{abstract}


\section{Introduction}
Real-time non-intrusive depth estimation of buried radioactive waste is often desirable due to the problems of generation of secondary wastes and increased dosage risks associated with traditional methods such as core sampling and logging. However, existing non-intrusive depth estimation methods such as \cite{Adams2012}, \cite{Shippen2011} are limited to depths of less than 3 cm or are based on empirical models. Therefore, this research presents a new non-intrusive depth estimation method based on an approximate three-dimensional (3D) linear attenuation model with improved depth estimation capability compared to the aforementioned methods.  

\section{Methodology}

\subsection{The approximate 3D linear attenuation model}
\begin{figure}[htbp]
\centerline{\includegraphics[width=7cm]{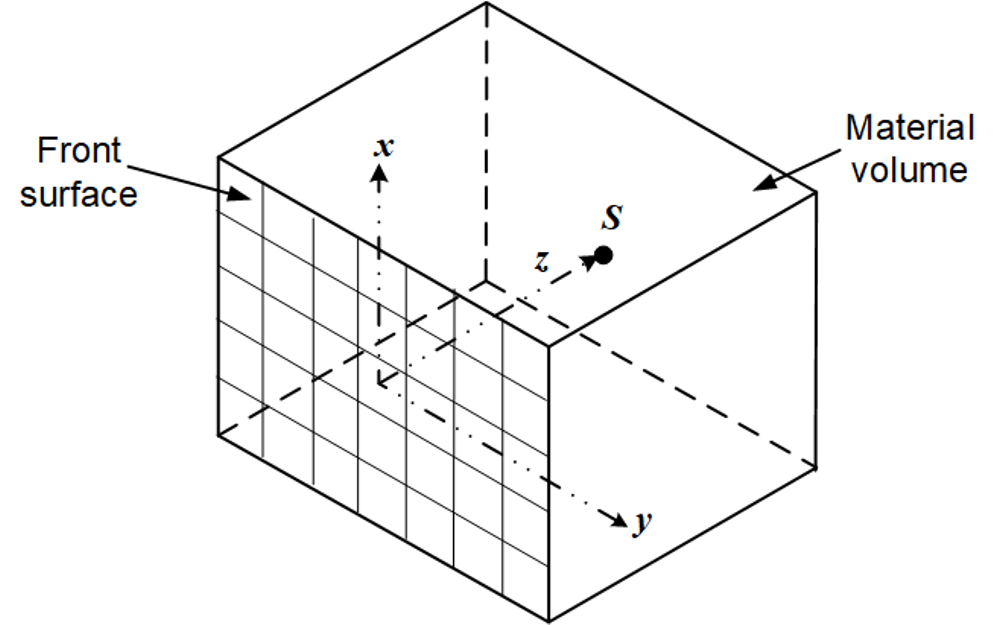}}
\caption{Radioactive source \textit{S} buried at depth \textit{z} inside a material volume}
\label{fig1}
\end{figure}
For a radioactive source \textit{S} buried at depth \textit{z} inside a material volume  as shown in Figure \ref{fig1}, the ratio of the intensity $I_{(x,y,z)}$ measured at any position on the surface of the material (i.e. \textit{x-y} plane) to that measured at (\textit{x,y}) = (0,0) is given by: 
\begin{equation}
log_e(J_{(x,y,z)}) \approx -\frac{\mu}{2z}(x^2+y^2)+log_e(K_{(x,y,0)})
\label{eq}
\end{equation}
where $J_{(x,y,z)}=\frac{I_{(x,y,z)}}{I_{(0,0,z)}}$, $\mu = \text{linear attenuation coefficient}$, and $K_{(x,y,0)}=\frac{I_{(x,y,0)}}{I_{(0,0,0)}}$. Equation \ref{eq} is the approximate 3D linear attenuation model. It can be observed that the depth of the radioactive source can be estimated from the gradient of the model. Furthermore, the gradient can be obtained by fitting a linear polynomial to the graph of the model obtained from the intensity measured at multiple position on the surface of the material volume. 

\subsection{Experiment}

\begin{figure}[htbp]
\centerline{\includegraphics[width=8cm]{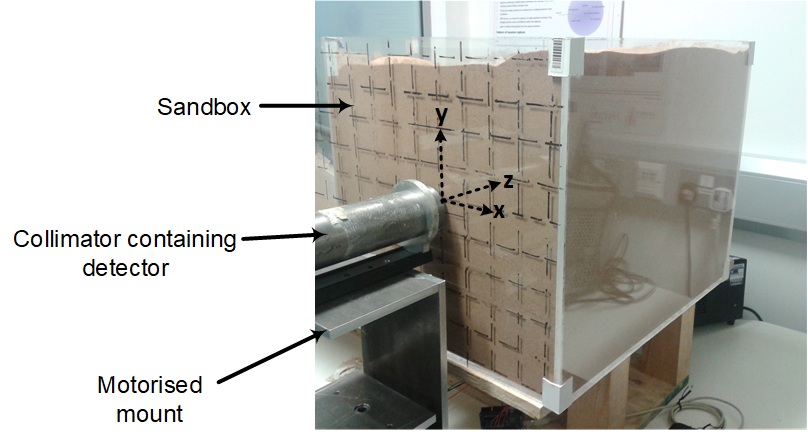}}
\caption{Experiment setup. The source was placed a varying positions along z-axis while the spectrum was measured at the grids marked on the surface of the sandbox}
\label{fig2}
\end{figure}

The setup for the experiment used to validate (\ref{eq}) is shown in Figure \ref{fig2}. It consists of a sandbox made of Perspex in which a sealed Cs-137 (329 kBq) point source was buried. The source was attached to one end of a graduated pipe whose other end protrudes behind the sandbox and was used to adjust the distance of the source from the scanning surface (i.e. along the \textit{z-axis}). The detector used in the experiment is an organic liquid scintillator (EJ-301) which was connected to a multichannel analyser. Furthermore, the detector was placed inside a hollow cylindrical tungsten collimator whose inner radius is approximately 4 cm. In addition, the surface of the sandbox was divided into 4 $\times$ 4 cm\textsuperscript{2} grids to correspond to the inner radius of the collimator. During the experiment, the position of the point source was varied along the \textit{z-axis} from 2 cm to 14 cm at 2 cm intervals. At each position, the spectrum of the buried source was acquired from 49 grids resulting in a total scan area of 28 $\times$ 28 cm\textsuperscript{2}.  Finally, the acquisition time was 10 minutes per grid.

\section{Results and Discussions}

\begin{figure}[htbp]
\centering
\includegraphics[width=9cm]{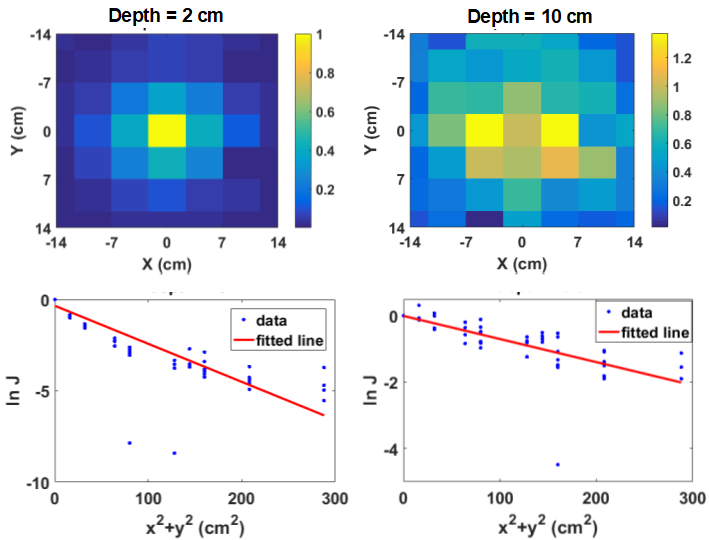}
\caption{Radiation images obtained at 2 cm and 10 cm depths (top row) and graphs of the model for both images respectively (bottom row)}
\label{fig3}
\end{figure}

Figure \ref{fig3} shows the normalised 2D radiation images of photons from the Compton peak region (top row) and corresponding graph of the model (bottom row) when the Cs-137 source was buried at depths of 2 cm and 10 cm respectively.  Photons from the Compton peak were used because the gamma spectrum from the EJ-301 detector is limited to the Compton continuum. It can be observed that the intensity of the source in the radiation images becomes increasingly defocussed as the depth increases. This is due to increasing scattering of the photons from the source by the sand matrix. Furthermore, as predicted by the model, a negative linear trend can be observed in the model graphs (i.e. Figure \ref{fig3} bottom row). Therefore, a linear polynomial was fitted to the data to estimate the depth. 

\begin{figure}[htbp]
\centerline{\includegraphics[width=6cm]{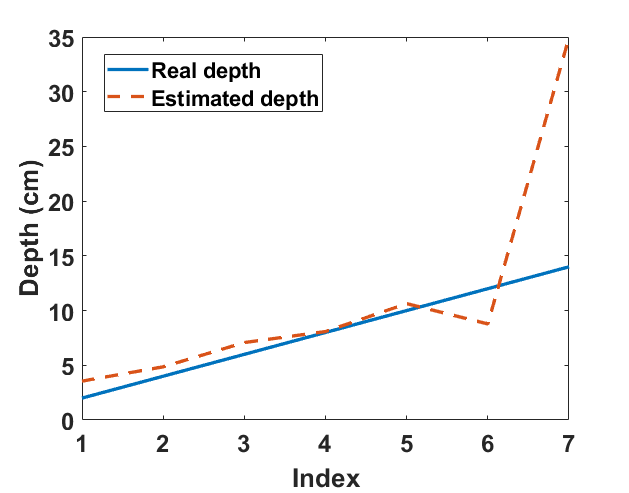}}
\caption{Real and estimated depths for the buried caesium-137 source.}
\label{fig4}
\end{figure}

\begin{figure}[htbp]
\centerline{\includegraphics[width=6cm]{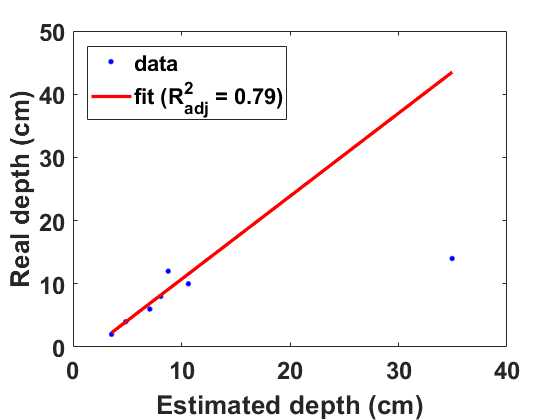}}
\caption{Linear fit for predicting the real depth from the estimated depth.}
\label{fig5}
\end{figure}

The estimated depths are shown in Figure \ref{fig4}. The linear attenuation coefficient of sand was calculated using mass coefficient values published in \cite{NationalInstituteofStandardsandTechnology2004} and elemental fraction values of sand obtained using Scanning Electron Microscopy. It can be observed from Figure \ref{fig4} that the estimated depths closely tracks the real depth up to 10 cm (i.e. index 5). However, a slight dip can be observed at 12 cm. This is likely due to experimental error. A sharp deviation of the estimated depth from the real depth can be observed beyond 12 cm. This is as a result of significant attenuation of the gamma rays resulting in large errors in the estimated depth. Finally, Figure \ref{fig5} shows that there is good linear fit between the real and estimated depth up to 12 cm which is a significant improvement compared to the techniques reported in \cite{Adams2012,Shippen2011}.

\section{Conclusion}
A new method for non-intrusive depth estimation of buried radioactive wastes have been presented. Experimental validation of the method showed that it is able to estimate depth of a Cs-137 point source buried up to 12 cm in sand. Therefore, the new method will be particularly useful for real-time characterisation of concrete and shallow underground contamination (e.g. leaks from underground pipelines used to transport liquid wastes) during decommissioning.

\end{document}